# ASYMPTOTIC QUANTUM SEARCH AND A QUANTUM ALGORITHM FOR CALCULATION OF A LOWER BOUND OF THE PROBABILITY OF FINDING A DIOPHANTINE EQUATION THAT ACCEPTS INTEGER SOLUTIONS

R. V. Ramos and J. L. de Oliveira

rubens@deti.ufc.br     jluzeilton@deti.ufc.br

*Department of Teleinformatic Engineering, Federal University of Ceara*

*Campus do Pici, C. P. 6007, 60455-740, Fortaleza, Brazil.*

Several mathematical problems can be modeled as a search in a database. An example is the problem of finding the minimum of a function. Quantum algorithms for solving this problem have been proposed and all of them use the quantum search algorithm as a subroutine and several intermediate measurements are realized. In this work, it is proposed a new quantum algorithm for finding the minimum of a function in which quantum search is not used as a subroutine and only one measurement is needed. This is also named asymptotic quantum search. As an example, we propose a quantum algorithm based on asymptotic quantum search and quantum counting able to calculate a lower bound of the probability of finding a Diophantine equation with integer solution.

## 1. Introduction

The Grover's quantum search algorithm is an important result in quantum computation that proves that quantum superposition can speed-up the task of finding a specific value within an unordered database. The quantum search is proved to use $O(N^{1/2})$ operations of the oracle (in comparison with the $O(N)$ operations of the best classical algorithm), indicating a quadratic speed-up [1-3]. Several mathematical problems can be modeled as a search, like the problem of finding the minimum (or the maximum) of a function. Thus, some algorithms for finding the minimum or maximum using quantum search have been proposed [4,5], using the generalization of the quantum search proposed in [6] as a subroutine that is called several times. Every time the quantum search is called, at the end a measurement is realized. The number of measurements in the algorithm proposed in [4] is $\Omega(\log^2 N)$, where $N$ is the number of elements in the database [7]. Aiming to reduce the number of measurements, Kowada at al [7] proposed a new quantum algorithm for finding the minimum of a function that realizes $O(\log N)$ measurements. Here, we go beyond, proposing a quantum algorithm for finding the minimum that realizes only one measurement, at the final of the algorithm. Furthermore, conversely the already proposed quantum algorithms for finding the minimum, the quantum search in the proposed algorithm is not used as a subroutine, indeed it is a part of the algorithm as any other quantum gate. In this work, as an example, we apply asymptotic quantum search together with quantum counting algorithm [3,8] in order to create a quantum algorithm able to calculate a lower bound of the probability of finding a Diophantine equation that accepts integer solutions. We are going to call this number by $\Omega_D$.

## 2. Quantum algorithm for finding the minimum of a function

Let us assume a function $f(x)=z$ where $x$ and $z$ are binary strings of $N$ bits, $f \{0,1\}^N \rightarrow \{0,1\}^N$. The goal is to determine the $x$ value that minimizes $f$. The circuit for the proposed quantum algorithm able to solve this problem is shown in Fig. 1.

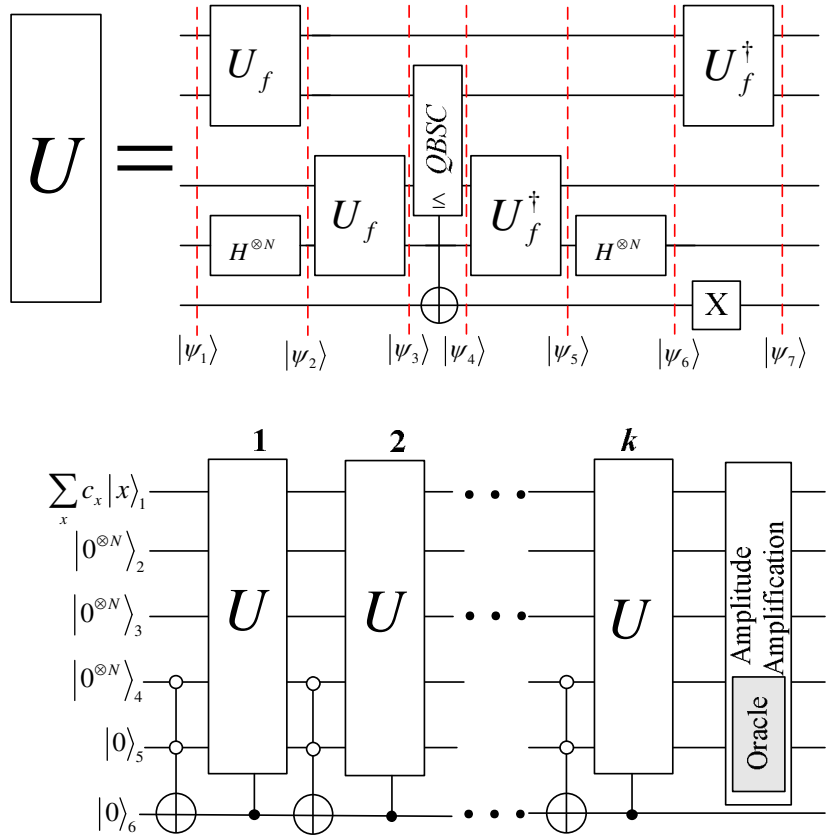

Fig. 1- Circuit for asymptotic quantum search.

In Fig. 1, $U_f$ is the quantum gate that implements the function $f$, $U_f|x\rangle|0\rangle=|x\rangle|f(x)\rangle$, the gate $QBSC$ is a quantum bit string comparator [9], $QBSC|x\rangle|y\rangle|0\rangle=|x\rangle|y\rangle|b\rangle$, where $b=0$ if $x>y$ and $b=1$ if $x\leq y$. At last, the oracle in the amplitude amplification recognizes if the quantum registers 4 and 5 are in the total state $|0^{\otimes N}\rangle|0\rangle$. In order to understand the quantum circuit in Fig. 1, we firstly show the operation of the quantum circuit $U$, following the states in the marked positions, as shown below:

$$|\psi_1\rangle = \left(\sum_x c_x |x\rangle_1 |0^{\otimes N}\rangle_2\right) |0^{\otimes N}\rangle_4 |0^{\otimes N}\rangle_3 |0\rangle_5 \tag{1}$$

$$|\psi_2\rangle = \left(\sum_x c_x U_f |x\rangle_1 |0^{\otimes N}\rangle_2\right) \left(H^{\otimes N} |0^{\otimes N}\rangle_4\right) |0^{\otimes N}\rangle_3 |0\rangle_5 = \left(\sum_x c_x |x\rangle_1 |f(x)\rangle_2\right) \left(\sum_y \frac{1}{\sqrt{2^N}} |y\rangle_4\right) |0^{\otimes N}\rangle_3 |0\rangle_5 \tag{2}$$

$$|\psi_3\rangle = \left(\sum_x c_x |x\rangle_1 |f(x)\rangle_2\right) \left(\sum_y \frac{1}{\sqrt{2^N}} U_f |y\rangle_4 |0^{\otimes N}\rangle_3\right) |0\rangle_5 = \left(\sum_x c_x |x\rangle_1 |f(x)\rangle_2\right) \left(\sum_y \frac{1}{\sqrt{2^N}} |y\rangle_4 |f(y)\rangle_3\right) |0\rangle_5 \tag{3}$$

$$|\psi_4\rangle = \sum_{\substack{x,y \\ f(x) \le f(y)}} \frac{c_x}{\sqrt{2^N}} |x\rangle_1 |f(x)\rangle_2 |y\rangle_4 |f(y)\rangle_3 |1\rangle_5 + \sum_{\substack{x,y \\ f(x) > f(y)}} \frac{c_x}{\sqrt{2^N}} |x\rangle_1 |f(x)\rangle_2 |y\rangle_4 |f(y)\rangle_3 |0\rangle_5 \tag{4.a}$$

$$|\psi_4\rangle = \left[\sum_x c_x |x\rangle_1 |f(x)\rangle_2 \sum_{j=1}^{n(x)} \frac{1}{\sqrt{2^N}} |y_j\rangle_4 |f(y_j)\rangle_3 \right] |1\rangle_5 +$$

$$\left[\sum_{\substack{x \\ x \ne x_{min}}} c_x |x\rangle_1 |f(x)\rangle_2 \sum_{j=1}^{2^N - n(x)} \frac{1}{\sqrt{2^N}} |y_j\rangle_4 |f(y_j)\rangle_3 \right] |0\rangle_5 \tag{4.b}$$

In (1)-(4.b), $N$ is the number of qubits, hence, there are $2^N$ elements in the database. In (4.b), $1 \le n(x) \le 2^N$ is the number of elements that obey $f(x) \le f(y)$ for a given $x$ and all $y$'s. Thus, the better the solution the larger is the value of $n(x)$. At the first term of (4.b), $y_j$ are ordered so that $f(y_j) \le f(y_{j+1})$ and $f(x) = f(y_1)$. At the second term of (4.b), $y_j$ are ordered so that $f(y_j) > f(y_{j+1})$ and $f(x) > f(y_1)$.

$$|\psi_5\rangle = \left[\sum_x c_x |x\rangle_1 |f(x)\rangle_2 \left(\sum_{j=1}^{n(x)} \frac{1}{\sqrt{2^N}} |y_j\rangle_4\right) |0^{\otimes N}\rangle_3 \right] |1\rangle_5 +$$

$$\left[\sum_{\substack{x \\ x \ne x_{min}}} c_x |x\rangle_1 |f(x)\rangle_2 \left(\sum_{j=1}^{2^N - n(x)} \frac{1}{\sqrt{2^N}} |y_j\rangle_4\right) |0^{\otimes N}\rangle_3 \right] |0\rangle_5 \tag{5}$$

$$|\psi_6\rangle = \left[c_{x_{min}} |x_{min}\rangle_1 |f(x_{min})\rangle_2 |0^{\otimes N}\rangle_4 |0^{\otimes N}\rangle_3 + \sum_{\substack{x \\ x \ne x_{min}}} c_x |x\rangle_1 |f(x)\rangle_2 \left(\sum_{j=1}^{n(x)} \frac{1}{\sqrt{2^N}} H^{\otimes N} |y_j\rangle_4\right) |0^{\otimes N}\rangle_3 \right] |1\rangle_5$$

$$+ \left[\sum_{\substack{x \\ x \ne x_{min}}} c_x |x\rangle_1 |f(x)\rangle_2 \left(\sum_{k=1}^{2^N - n(x)} \frac{1}{\sqrt{2^N}} H^{\otimes N} |y_k\rangle_4\right) |0^{\otimes N}\rangle_3 \right] |0\rangle_5 \tag{6.a}$$

$$|\psi_6\rangle = \left[ c_{x_{min}} |x_{min}\rangle_1 |f(x_{min})\rangle_2 |0^{\otimes N}\rangle_4 |0^{\otimes N}\rangle_3 + \sum_{\substack{x \\ x \neq x_{min}}} c_x n(x) 2^{-N} |x\rangle_1 |f(x)\rangle_2 |0^{\otimes N}\rangle_4 |0^{\otimes N}\rangle_3 \right] |1\rangle_5$$

$$+ \left[ \sum_{\substack{x \\ x \neq x_{min}}} c_x |x\rangle_1 |0^{\otimes N}\rangle_2 \left( \sum_{\substack{j=1 \\ H^{\otimes N}|y_j\rangle \neq |0^{\otimes N}\rangle}}^{n(x)} \frac{1}{\sqrt{2^N}} H^{\otimes N} |y_j\rangle_4 \right) |0^{\otimes N}\rangle_3 \right] |1\rangle_5 \quad (6.b)$$

$$+ \left[ \sum_{\substack{x \\ x \neq x_{min}}} c_x |x\rangle_1 |f(x)\rangle_2 \left( \sum_{k=1}^{2^N - n(x)} \frac{1}{\sqrt{2^N}} H^{\otimes N} |y_k\rangle_4 \right) |0^{\otimes N}\rangle_3 \right] |0\rangle_5$$

Here, the expression $H^{\otimes N}|y_j\rangle \neq |0^{\otimes N}\rangle$ means the term $|0^{\otimes N}\rangle$ in the result of $H^{\otimes N}|y_j\rangle$ was already taken into account in the first bracket at (6.b). Finally

$$|\psi_7\rangle = U\left[ \left( \sum_x c_x |x\rangle_1 |0^{\otimes N}\rangle_2 \right) |0^{\otimes N}\rangle_4 |0^{\otimes N}\rangle_3 |0\rangle_5 \right] = \left[ \sum_x c_x n(x) 2^{-N} |x\rangle_1 |0^{\otimes N}\rangle_2 |0^{\otimes N}\rangle_4 |0^{\otimes N}\rangle_3 \right] |0\rangle_5$$

$$+ \left[ \sum_{\substack{x \\ x \neq x_{min}}} c_x |x\rangle_1 |0^{\otimes N}\rangle_2 \left( \sum_{\substack{j=1 \\ H^{\otimes N}|y_j\rangle \neq 0^{\otimes N}}}^{n(x)} \frac{1}{\sqrt{2^N}} H^{\otimes N} |y_j\rangle_4 \right) |0^{\otimes N}\rangle_3 \right] |0\rangle_5 \quad (7)$$

$$+ \left[ \sum_{\substack{x \\ x \neq x_{min}}} c_x |x\rangle_1 |0^{\otimes N}\rangle_2 \left( \sum_{k=1}^{2^N - n(x)} \frac{1}{\sqrt{2^N}} H^{\otimes N} |y_k\rangle_4 \right) |0^{\otimes N}\rangle_3 \right] |1\rangle_5$$

The worse the solution the lower is the value of *n(x)* and the faster is the decay. Now, using *k*-times the gate *U* together with the multi-controlled CNOTs, the quantum state just before the quantum amplification is:

$$\left[ \sum_x c_x \left[ n(x) 2^{-N} \right]^k |x\rangle_1 |0^{\otimes N}\rangle_2 |0^{\otimes N}\rangle_4 |0^{\otimes N}\rangle_3 \right] |0\rangle_5$$

$$+ \sum_{r=1}^{k} \left[ \sum_{\substack{x \\ x \neq x_{min}}} c_x \left[ n(x) 2^{-N} \right]^{r-1} |x\rangle_1 |0^{\otimes N}\rangle_2 \left( \sum_{\substack{j=1 \\ H^{\otimes N}|y_j\rangle \neq |0^{\otimes N}\rangle}}^{n(x)} \frac{1}{\sqrt{2^N}} H^{\otimes N} |y_j\rangle_4 \right) |0^{\otimes N}\rangle_3 \right] |0\rangle_5 \quad (8)$$

$$+ \sum_{r=1}^{k} \left[ \sum_{\substack{x \\ x \neq x_{min}}} c_x \left[ n(x) 2^{-N} \right]^{r-1} |x\rangle_1 |0^{\otimes N}\rangle_2 \left( \sum_{k=1}^{2^N - n(x)} \frac{1}{\sqrt{2^N}} H^{\otimes N} |y_k\rangle_4 \right) |0^{\otimes N}\rangle_3 \right] |1\rangle_5$$

At the amplitude amplification, the oracle recognizes the state $|0^{\otimes N}\rangle|0\rangle$ in the fourth and fifth quantum registers. Therefore, only the first term in (8) will have their amplitudes amplified. Looking closer the first term in (8), one sees that the amplitude of the searched answer is $c_{x_{min}}$, since $n(x)=2^N$ for $x=x_{min}$. The second term with largest amplitude has $n(x)=2^N-1$, hence, after the $k$-th application of $U$ its amplitude will be $c_x(1-2^{-N})^k$. Thus, if $k$ is large enough only the term corresponding to the searched answer will have considerable amplitude and, after amplitude amplification, the searched answer will be obtained with high probability in a single set of measurements. For example, choosing $c_x = 1/\sqrt{2^N}$ for all $x$, the largest amplitude after $k$ usage of $U$ (before amplitude amplification) is $c_{x_{min}} = 1/\sqrt{2^N}$ while the second largest amplitude will be $c_x = c_{x_{min}}(1-2^{-N})^k$. The oracle in the amplitude amplification is applied $\pi/(4\theta)$ times, where [9]

$$\sin(\theta_k) = \sqrt{p_{good}}. \qquad (9)$$

If $k$ is large enough, then $p_{good} \approx 2^{-N}$.

In order to make the analysis of the complexity of the proposed algorithm let us count how many application of $U$ in Fig. 1 are necessary in order to let the second largest amplitude inside the first bracket of (8) very close to zero, considering that the initial state is a equally weighted superposition, $c_x=2^{-N/2}$ for all $x$. A numerical calculation shows that $(1-2^{-N})^k \leq 3.355 \cdot 10^{-4}$ for $k=2^{N+3}$ and $N<54$. Hence, the number of times that $U$ is used is if the same order of the size of the database. Once the last application of $U$ was realized the amplitude amplification is performed and the correct answer will be obtained, in a single measurement, with probability close to one, after $O(2^{N/2})$ queries to the amplitude amplification's oracle. The comparison of the asymptotic quantum search (AQS) here proposed with the quantum algorithms LM proposed in [7] and DH proposed in [4] can be seen in Table 1.

|        | $N_1$     | $N_2$        | $N_3$            | Probability of success |
|--------|-----------|--------------|------------------|------------------------|
| AQS    | $2^{N+3}$ | $O(2^{N/2})$ | 1                | $1-O(2^{-N})$          |
| LM [7] | 0         | $O(2^{N/2})$ | $O(\log(2^N))$   | at least ½             |
| DH [4] | 0         | $O(2^{N/2})$ | $O(\log^2(2^N))$ | at least ½             |

Table 1 – Comparison between the quantum algorithms AQS, LM [7] and DH [4]. $N_1$ is the number of operations realized before the quantum search, $N_2$ is the number of queries to the oracle and $N_3$ is the number of measurements performed.

Observing Table 1, we can see that the price to be paid in order to have a high probability of success (close to 1) and only one measurement is the number of operations realized before the usage of the amplitude amplification.

## 3. Quantum algorithm for finding a lower bound of $\Omega_D$

A Diophantine equation is a polynomial equation with any number of unknowns and with integer coefficients. For example, $(x+1)^n+(y+2)^n+(z+3)^n-Cxyz=0$, where $C$ and $n \in Z$. Since there is not a universal process to determine whether any Diophantine equation accepts or not integer solutions, the best that one can do is to use a brute-force method. In this case, one can not determine the probability of finding a Diophantine equation that accepts integer solutions, $\Omega_D$, exactly, but it is possible to determine a lower bound for it. In this direction, the quantum algorithm proposed can be used to implement a brute-force attack (using the intrinsic quantum parallelism) and the accuracy of the answer obtained is limited by the amount of qubits used, the larger the amount of qubits used the closer to $\Omega_D$ is the obtained answer. Since only a finite amount of qubits can be used, the answer obtained is simply a lower bound of $\Omega_D$. Since the (infinite) set of Diophantine equations considered is enumerable, we will use integer numbers of $N$ bits to enumerate them. The quantum circuit for the proposed quantum algorithm is shown in Fig. 2. In this quantum circuit, we assume that the set of Diophantine equations considered has only three unknowns $(x,y,z)$, however, the extension to a larger number of unknowns is straightforward.

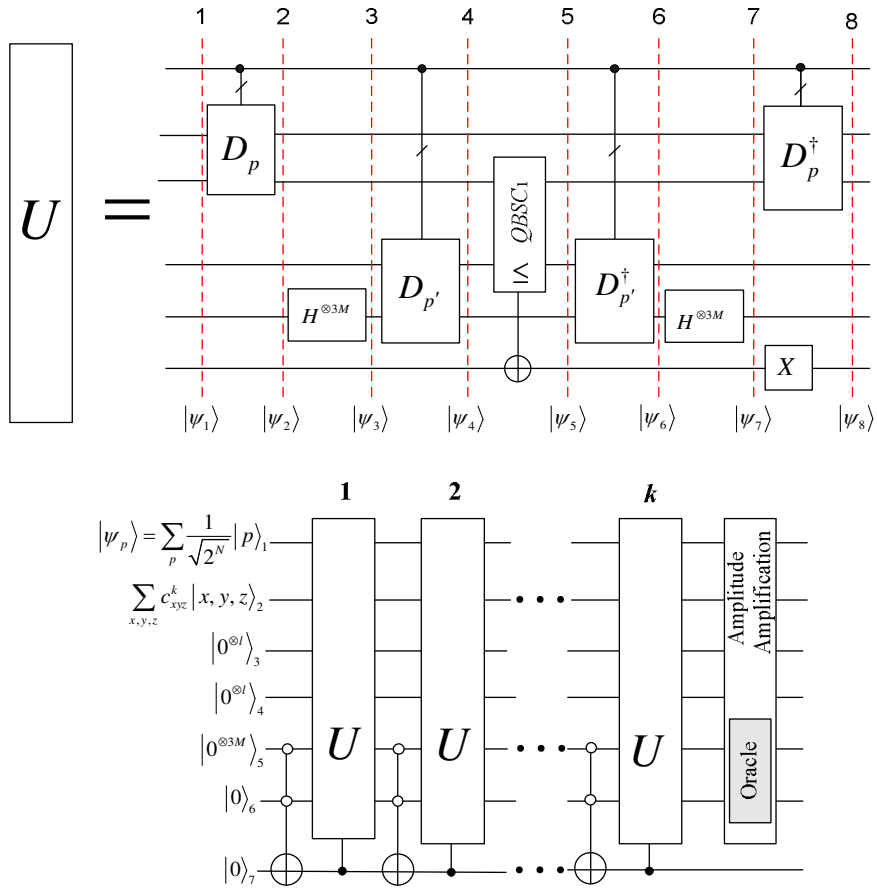

Fig. 2- Quantum circuit used to calculate a lower bound for $\Omega_D$ (the quantum counting part is absent).

In Fig. 2, $D_p$ is the quantum gate that calculates the absolute value of the Diophantine equation number $p$. The $QBSC_1$ works as explained in Section 2. At last, the oracle in the amplitude amplification recognizes if the registers 5 and 6 in Fig. 2 are in the total state $|0^{\otimes 3M}\rangle|0\rangle$. The circuit in Fig. 2 tests, simultaneously, $2^N$ different Diophantine equations. Each unknown is represented by an $M$-bit string while the result of the Diophantine equation $|D_p(x,y,z)|$ is coded in an $l$-bit string. Basically, the quantum circuit in Fig. 2 is an asymptotic quantum search that finds the minimum absolute value of each Diophantine equation when the input variables vary from 0 to $2^M-1$. Thus, the strategy is: to building a superposition over all equations and variables, amplifying the amplitude of the variables that minimizes $|D_p(x,y,z)|$ and, then, counting the number of equations for which such minimum is zero. In order to understand the quantum circuit in Fig. 2, we firstly show the operation of the quantum circuit $U$, following the states in the marked positions, as shown below:

$$|\psi_1\rangle = \sum_p \frac{1}{\sqrt{2^N}} |p\rangle_1 \left( \sum_{x,y,z} c_{xyz} |x,y,z\rangle_2 |0^{\otimes l}\rangle_3 \right) \left( |0^{\otimes 3M}\rangle_5 |0^{\otimes l}\rangle_4 \right) |0\rangle_6 \tag{10}$$

$$|\psi_2\rangle = \sum_p \frac{1}{\sqrt{2^N}} |p\rangle_1 \left( \sum_{x,y,z} c_{xyz} |x,y,z\rangle_2 ||D_p(x,y,z)|\rangle_3 \right) \left( |0^{\otimes 3M}\rangle_5 |0^{\otimes l}\rangle_4 \right) |0\rangle_6 \tag{11}$$

$$|\psi_3\rangle = \sum_p \frac{1}{\sqrt{2^N}} |p\rangle_1 \left( \sum_{x,y,z} c_{xyz} |x,y,z\rangle_2 ||D_p(x,y,z)|\rangle_3 \right) \left( \sum_{x',y',z'} \frac{1}{\left(\sqrt{2^M}\right)^3} |x',y',z'\rangle_5 |0^{\otimes l}\rangle_4 \right) |0\rangle_6 \tag{12}$$

$$|\psi_4\rangle = \sum_p \frac{1}{\sqrt{2^N}} |p\rangle_1 \left( \sum_{x,y,z} c_{xyz} |x,y,z\rangle_2 ||D_p(x,y,z)|\rangle_3 \right) \left( \sum_{x',y',z'} \frac{1}{\left(\sqrt{2^M}\right)^3} |x',y',z'\rangle_5 ||D_p(x',y',z')|\rangle_4 \right) |0\rangle_6 \tag{13}$$

$$|\psi_5\rangle = \sum_p \frac{1}{\sqrt{2^N}} |p\rangle_1 \sum_{x,y,z} c_{xyz} |x,y,z\rangle_2 ||D_p(x,y,z)|\rangle_3 \sum_{\substack{i,j,m \\ D_p(x,y,z) \leq D_p(x'_i, y'_j, z'_m)}} \frac{1}{\left(\sqrt{2^M}\right)^3} |x'_i, y'_j, z'_m\rangle_5 ||D_p(x'_i, y'_j, z'_m)|\rangle_4 |1\rangle_6 + |\phi_1\rangle \tag{14}$$

$$|\psi_6\rangle = \sum_p \frac{1}{\sqrt{2^N}} |p\rangle_1 \sum_{x,y,z} c_{xyz} |x,y,z\rangle_2 ||D_p(x,y,z)|\rangle_3 \sum_{\substack{i,j,m \\ D_p(x,y,z) \leq D_p(x'_i, y'_j, z'_m)}} \frac{1}{\left(\sqrt{2^M}\right)^3} |x'_i, y'_j, z'_m\rangle_5 |0^{\otimes l}\rangle_4 |1\rangle_6 + |\phi_2\rangle \tag{15}$$

$$|\psi_7\rangle = \sum_p \frac{1}{\sqrt{2^N}} |p\rangle_1 \left( \begin{array}{c} \sum_{s=1}^{t_p} c_{xyz^p_{opt_s}} |x^p_{opt_s}, y^p_{opt_s}, z^p_{opt_s}\rangle_2 ||D_p(x^p_{opt_s}, y^p_{opt_s}, z^p_{opt_s})|\rangle_3 |0^{\otimes 3M}\rangle_5 + \\ \sum_{\substack{x,y,z \\ x,y,z \neq x^p_{opt}, y^p_{opt}, z^p_{opt}}} c_{xyz} n(x,y,z) 2^{-3M} |x,y,z\rangle_2 ||D_p(x,y,z)|\rangle_3 |0^{\otimes 3M}\rangle_5 + |\phi_3\rangle \end{array} \right) |0^{\otimes l}\rangle_4 |1\rangle_6 + |\phi_4\rangle \tag{16}$$

Finally

$$|\psi_8\rangle = U\sum_p \frac{1}{\sqrt{2^N}}|p\rangle_1 \left(\sum_{x,y,z} c_{xyz}|x,y,z\rangle_2 |0^{\otimes l}\rangle_3\right)\left(|0^{\otimes 3M}\rangle_5 |0^{\otimes l}\rangle_4\right)|0\rangle_6 =$$

$$\sum_p \frac{1}{\sqrt{2^N}}|p\rangle_1 \left(\sum_{s=1}^{t_p} c_{xyz^p_{opt_s}} \left|x^p_{opt_s}, y^p_{opt_s}, z^p_{opt_s}\right\rangle_2 + \sum_{\substack{x,y,z \\ x,y,z \neq x^p_{opt}, y^p_{opt}, z^p_{opt}}} c_{xyz} n(x,y,z) 2^{-3M}|x,y,z\rangle_2\right)|0^{\otimes l}\rangle_3 |0^{\otimes 3M}\rangle_5 |0^{\otimes l}\rangle_4 |0\rangle_6 + |\phi\rangle \qquad (17)$$

In (17), $1 \leq n(x,y,z) \leq 2^{3M}$ is the number of elements that obey $|D_p(x,y,z)| \leq |D_p(x',y',z')|$ for a given $x,y,z$ and all $x',y',z'$. Thus, the better the solution the larger is the value of $n(x,y,z)$. We assume that, for the Diophantine equation number $p$ there are $t_p$ minimums. Moreover $n\left(x^p_{opt_s}, y^p_{opt_s}, z^p_{opt_s}\right) = 2^{3M}$ for $s=1,2,\ldots,t_p$. The states $|\phi_i\rangle$ $i=1,2,3,4$ and $|\phi\rangle$ are states with undesired terms. Now, using $k$-times the gate $U$ together with the multi-controlled CNOT gates, the quantum state just before the quantum amplification is:

$$\sum_p \frac{1}{\sqrt{2^N}}|p\rangle_1 \left(\sum_{s=1}^{t_p} c_{xyz^p_{opt_s}} \left|x^p_{opt_s}, y^p_{opt_s}, z^p_{opt_s}\right\rangle_2 + \sum_{\substack{x,y,z \\ x,y,z \neq x^p_{opt}, y^p_{opt}, z^p_{opt}}} c_{xyz}\left[n(x,y,z)2^{-3M}\right]^k |x,y,z\rangle_2\right)|0^{\otimes l}\rangle_3 |0^{\otimes 3M}\rangle_5 |0^{\otimes l}\rangle_4 |0\rangle_6 + |\Phi\rangle \qquad (18)$$

The state $|\Phi\rangle$ represents the uninteresting terms. At the amplitude amplification, the oracle recognizes the state $|0^{\otimes 3M}\rangle|0\rangle$ in the fifth and sixth quantum registers. Therefore, only the first term in (18) will have their amplitudes amplified. Looking closer the first term in (18), one sees that the amplitude of one of the searched answers is $c_{xyz^p_{opt_s}}$. The second term with largest amplitude has $n(x,y,z)=2^{3M}-1$, hence, after the $k$-th application of $U$ its amplitude will be $c_{xyz}(1-2^{-3M})^k$. Thus, if $k$ is large enough only the term corresponding to the searched answer will have considerable amplitude and, after amplitude amplification, it will be obtained with high probability in a single measurement. For example, choosing $c_{xyz} = 1/\sqrt{2^{3M}}$ for all $xyz$, the largest amplitude after $k$ usage of $U$ (before amplitude amplification) is $c_{xyz^p_{opt_s}} = 1/\sqrt{2^{3M}}$ while the second largest amplitude will be $c_{xyz} = c_{xyz^p_{opt_s}}\left(1-2^{-3M}\right)^k$. Once more The oracle in the amplitude amplification is applied $\pi/(4\theta)$ times, where $\sin(\theta_k) = \sqrt{p_{good}}$. If $k$ is large enough, then $p_{good} \approx t_p 2^{-3M}$. Thus, the final state after the amplitude amplification is

$$|\psi_{end}\rangle \approx \sum_p \frac{1}{\sqrt{2^N}}|p\rangle \sum_{s=1}^{t_p} \frac{1}{\sqrt{t_p}} \left|x^p_{opt_s}, y^p_{opt_s}, z^p_{opt_s}\right\rangle \left\| D_p\left(x^p_{opt_s}, y^p_{opt_s}, z^p_{opt_s}\right)\right\| \rangle + |err\rangle \qquad (19)$$

In (19) one can see that, for each one of the $2^N$ Diophantine equations, the algorithm calculates the minimums values testing $2^M$ different values for each unknown. Moreover, the term $|err\rangle$ represents the remained unwanted states due to the non ideal amplitude amplification.

In order to calculate the lower bound of $\Omega_D$, we use the state (19) as input of a quantum counting algorithm whose oracle recognizes a zero in the last register of (19), that is, $\left|D_p\left(x_{opt}^p, y_{opt}^p, z_{opt}^p\right)\right| = 0$. The quantum counting algorithm returns the number of elements in the database that pass in the oracle test. Let us name this value by $r(N,M)$. Hence,

$$\Omega_D = \lim_{N,M \to \infty} r(N,M)/2^N \tag{20}$$

and a lower bound for $\Omega_D$ is simply $\Omega_D^{est} = r(N,M)/2^N$.

The complexity of this algorithm is based on the complexity of the asymptotic quantum search given in Section 2. In order to prepare the state for quantum search (equation (18)) $2^{(3M+3)}$ usages of the gate $U$ are necessaries (decreasing the factor of the second largest amplitude by $10^{-4}$). The complexity of the amplitude amplification is $O(2^{3M/2})$ and the complexity of the quantum counting is $O(2^{N/2})$. At last, only one measurement is required at the end of the quantum counting.

## 4. Discussions

The problem to decide if a given Diophantine equation has or not integer solutions is, according to Church-Turing computability thesis, an undecidable problem. In fact, this problem is related to the halting problem for Turing machine: The Diophantine equation problem could be solved if and only if the Turing halting problem could be [10]. Since there is not a universal process to decide if a given program will halt or not a Turing machine, the best that one can do is to run the program and wait for a halt. The point is: one can never know if he/she waited enough time in order to observe the halt. The Diophantine equation problem has a similar statement. Since there is not a universal procedure to decide if a given Diophantine equation accepts an integer solution, the best that one can do in order to check if it accepts or not an integer solution, is to test the integer numbers. The point here is: one can never know the answer because there are infinite integers to be tested. Associated to the halting problem is the Chaitin number $\Omega$, the probability of a given program to halt the Turing machine [11-17]. The number $\Omega$ is an incompressible number and it can not be calculated. Similarly, one can define the probability of finding a Diophantine equation with integer solutions, $\Omega_D$. Obviously, $\Omega_D$ is also a number that can not be calculated, because the knowledge of $\Omega_D$ implies in the knowledge of a way to decide if any given Diophantine equation accepts or not any integer solution that, by its turn, implies in the knowledge of $\Omega$ and in the solution of the halting problem. As happens with $\Omega$, it is possible to calculate classically a lower bound for $\Omega_D$, however, it must be a brute-force algorithm. Classical algorithms for calculation of $\Omega_D$ will differ basically in the method that the zeros of the Diophantine equation are obtained. The quantum algorithm here proposed is also a brute-force algorithm but its advantage is the quantum parallelism, since several Diophantine equations are tested simultaneously.

## 4. Conclusions

It was proposed a new quantum algorithm for finding the minimum of a function that requires only one measurement. This is important since the complete circuit is simplified and measurements of qubits are not noise free. This algorithm is an asymptotic quantum search. Its advantages are the high probability to get the right result and the fact that quantum search and measurements are realized only once. The disadvantage is the number of operations to be realized before the quantum search. As an example of its application, we constructed a quantum algorithm, employing quantum counting, for finding a lower bound for the probability of finding a Diophantine equation that accepts integer solutions.

## Acknowledge

The authors thank the anonymous reviewer for useful comments.